# Global Epigenetic State Network Governs Cellular Pluripotent Reprogramming and Transdifferentiation

Ping Wang[1]*, Chaoming Song[2,3]*[§], Hang Zhang[1], Zhanghan Wu[1,4], Jianhua Xing[1][§]

[1]Department of Biological Sciences, Virginia Tech, Blacksburg, VA 24060, USA [2]Center for Complex Network Research, Departments of Physics, Biology, and Computer Science, Northeastern University, Boston, MA 02115, USA.

[3]Department of Medicine, Harvard Medical School, and Center for Cancer Systems Biology, Dana-Farber Cancer Institute, Boston, MA 02115, USA.

[4]National Heart, Lung and Blood Institutes, National Institutes of Health, Bethesda, MD 20892

*These authors contributed equally to this work

[§]To whom correspondence should be addressed. E-mail: jxing@vt.edu, c.song@neu.edu

**How do mammalian cells that share the same genome exist in notably distinct phenotypes, exhibiting differences in morphology, gene expression patterns, and epigenetic chromatin statuses? Furthermore how do cells of different phenotypes differentiate reproducibly from a single fertilized egg? These fundamental questions are closely related to a deeply rooted paradigm in developmental biology that cell differentiation is irreversible. Yet, recently a growing body of research suggests the possibility of cell reprogramming, which offers the potential for us to convert one type of cell into another [1-5]. Despite the significance of quantitative understandings of cell reprogramming [6], theoretical efforts often suffer from the complexity of large circuits maintaining cell phenotypes coupled at many different epigenetic and gene regulation levels. To capture the global architecture of cell phenotypes, we propose an "epigenetic state network" approach that translates the classical concept of an epigenetic landscape into a simple-yet-predictive mathematical model. As a testing case, we apply the approach to the reprogramming of fibroblasts (FB) to cardiomyocytes (CM). The epigenetic state network for this case predicts three major pathways of reprogramming. One pathway goes by way of induced pluripotent stem cells (iPSC) and continues on to the normal pathway of cardiomyocyte differentiation. The other two pathways involve transdifferentiation (TD) either indirectly through cardiac progenitor (CP) cells or directly from fibroblast to cardiomyocyte. Numerous experimental observations support the predicted states and pathways.**

One theoretical approach towards modelling cell reprogramming relies on a set of coupled differential equations, offering a quantitative description of gene regulation dynamics[7-9]. Alternatively, Waddington suggested an epigenetic landscape picture[10] of establishing cell fates during development that resembles a ball rolling down a "Landscape" to the point of lowest local elevation [11], where the axes and basins represent molecular concentrations and stable cell phenotypes, respectively [8,12-14]. The Waddington's epigenetic landscape is analogous to the energy landscape concept widely used in chemistry and physics studies, e.g. in protein physics each basin represents one stable protein conformation [15]. The emphases of these two approaches, however, are sharply different. The ODE model integrates molecular biology details piece-by-piece, from which the system dynamics can be deduced quantitatively. Nevertheless, the involvement of a large number of genes and lineages during cell reprogramming has led to increasing awareness that a global view of the system dynamics is necessary. The quantitative and detailed description of a high-dimensional dynamic system provided by an ODE model, ironically, makes it challenging to generate a global view. In contrast, the popular Waddington's landscape concept provides an intuitive and transparent birds' eye view, yet may only serve as a phenomenological metaphor lacking of either mechanistic details or quantitatively predictive power [16]. Since the last decade, many efforts have been made towards quantifying the epigenetic landscape [16,17], yet systematic application and analysis beyond simple model systems is still missing. In particular, generalization of rigorous descriptions [17] to higher dimensions is challenging (SI Section 1). Here we present a novel approach that allows us to construct epigenetic landscapes quantitatively from ODE models, offering a unified framework capturing the reprogramming dynamics at large scales.

The long term behaviour of a dynamic system is largely determined by its attractors [18]. For cell reprogramming, we are particularly interested in the fixed-point attractors that

4represent different cell phenotypes. The fact that the fixed-point attractors are stable under infinitesimal perturbation implies that the cell phenotypes are robust most of the time. However, stochastic noise provides an unbounded perturbative force that drives the cell from one state to others [16]. The pair of most probable transition paths between two neighbouring attractors in general pass around a first-order saddle point (fixed-point with only one component unstable) [17,19,20]. Therefore, the regulation dynamics described with ODE equations can be approximated by a stochastic walk on a weighted network (Epigenetic State Network or ESN) whose nodes correspond to the fixed-point attractors, and edges (*i.e.*, links between nodes) represent the first-order saddle points connecting two neighbouring attractors with the weight associated with each edge representing the transition rate.Figure 1 demonstrates the ESN approach applied for a two-gene ($x_1$, $x_2$) regulatory circuit, a simple prototype model sustaining multiple cell reprogramming paths [8,12]. The vector field in Fig. 1A and the associated ESN in Fig. 1B reveal three paths: the first two occur through a progenitor with both $x_1$ and $x_2$ high then differentiating into another phenotype ($x_1$ high and $x_2$ low, or $x_1$ low and $x_2$ high), and another is a direct path connecting the two phenotypes without going through the progenitor. Furthermore, Fig. 1C, Table S2 and S3 show that both the number of fixed-point attractors and network topology change with different kinetic parameters. In this case, ESN provides an alternative representation of cell differentiation and reprogramming that captures the major dynamics of the underlying two dimensional vector fields (see Fig. 1A), consistent with traditional approaches such as bifurcation analysis [18]. However, the vector field of a large system is not directly visible and the corresponding lower dimensional profile offers only partial information of the underlying landscape. The ESN approach, however, coarse grains the essential characters of an arbitrary dimensional dynamic system into a low-dimensional network graph and thus can reveal the global features transparently.

To demonstrate the practical power of our approach, we apply the ESN analysis to the FB-iPSC-CM system. This system has received much attention within the last few years [1-5], and experiments have shown FB to CM reprogramming through iPSC, as well as direct/indirect transdifferentiation (Fig. 2A). This system can be modelled by a 10-master gene regulatory circuit (Fig. 2B) based on the IPA® (@Ingenuity) database (SI Section 2). The chromatin status determining the accessibility of associated regulators is the key factor that controls different cell phenotypes [21]. We are particularly interested in the reprogramming dynamics under three different "plastic" chromatin statuses [21] (see SI Section 3 and Fig. S3): 1) fibroblast-iPSC pluripotent reprogramming ESN (PR-ESN), for which only the FB and stem cell (SC) regions are accessible; 2) fibroblast-cardiomyocyte transdifferentiation ESN (TD-ESN), for which only the FB and CM regions are accessible; 3) globally open ESN (GO-ESN), for the embryonic stem cells and some pathological cases such as cancer cells, the chromosomes are highly decondensed and the regulatory network may be largely accessible [21,22]. The non-genetic cellular heterogeneity indicates an inherent variation of global gene expression patterns among the cells [23]. Therefore, the ESN analysis is applied, over an ensemble of parameters through Monte Carlo sampling reflecting this cell-to-cell variation., instead of one fixed set of kinetic constants,

Figure 2C-E show the PR-ESN, TD-ESN and GO-ESN averaged over $10^5$ Monte Carlo realizations (SI Section 3, Fig. S4 & S5), where the node and edge sizes are proportional to their occurrence probabilities. We find that a large amount of cell states in PR-ESN (Fig. 2C) share similar expression patterns, implying the non-genetic heterogeneity at the single cell level. The PR-ESN consists of pathways connecting mostly the FB and the stem cell states (iPSC1 and iPSC2), dominated by a major path through a pre-iPSC (PP) state. The TD-ESN instead, shows major paths between the FB and CM states passing through the PP state, intermediate states with CM progenitor regulators on, and FC state (an intermediate state with both FB and CM regulators on) and other (Fig. 2D). The GO-ESN (Fig. 2E) is approximately a combination of the PR-ESN and TD-ESN (Fig. 2CD), suggesting that the ESNs under two different chromatin statuses (PR-ESN and TD-ESN) are largely uncorrelated with each other. Counterintuitively, despite that the FB and CM transcription factors interact indirectly through progenitor regulators (Fig. S2), both TD-ESN and GO-ESN show a pathway connecting the FB and CM states through the FC state without passing through the upstream PP state, and is confirmed by empirical observations of DTD [5], indicating that the system dynamics is best captured by the topology of the ESN rather the original regulatory circuit.

To investigate the reprogramming dynamics quantitatively, we measure for all three ESNs the transition rates for each pair of neighbouring states. The fact that the transition rate distribution is highly uneven indicates that the reprogramming dynamics is mostly dominated by those paths with large transition rates (or equivalently, small passage times) (Fig. S6). Therefore, we calculate the minimum-spanning tree (MST) that captures the 'backbone' of the ESN by optimizing the average passage time (defined as the reciprocal of the transition rate, see SI 3). Figure 3A-C show the MSTs calculated for PR-ESN, TD-ESN and GO-ESN respectively, revealing star-like structures that include the most important reprogramming pathways. Remarkably, states (nodes) in each ESN organize into clusters with a major state surrounded by less probable states with a similar expression pattern. The MSTs of the PR-ESN and TD-ESN represent subgraphs of the GO-ESN-MST, reflecting again the fact that the PR-ESN and TD-ESN are both independent subsystems of GO-ESN. We calculate the relative expression level of each cluster by averaging the expressions over the states within the same cluster (Figs. 3DEF), finding that the regulator expressions of different cell phenotypes are largely exclusive to each other (e.g. fibroblast and stem cell). This reflects the mutual inhibitions of regulators representing different cell phenotypes.

To facilitate comparison with experiments, we examine the published microarray data in Refs 24-26: For pluripotent reprogramming, experiments observed that the dynamical evolution from mouse embryonic fibroblast (MEF, fibroblast regulators turning on, all regulators turning off), to pre-iPSC (PP, silencing fibroblast regulators without expressing SC regulators) then mouse iPSC (MiPSC, SC regulators turning on) states sequentially (Fig. 4A)[27], , confirmed the prediction from the PR-ESN (Fig. 3D). To examine the predicted state heterogeneity, we take the SC states (red color nodes in Fig. 3A-C) for example, and find that the expression patterns observed from different experiments do indeed slightly differ from each other, as shown in Fig. 4B for the mouse embryonic stem cell (MESC), the ventricular cardiomyocyte induced pluripotent stem cell (ViPSC), and the mouse tail-tip fibroblast induced pluripotent stem cell (TiPSC). Figure. 4C reports the gene expression patterns measured in the differentiation experiments from human induced pluripotent stem cell (HiPSC) into cardiomyocyte (CM), which are consistent with our predictions as well. For transdifferentiation (Fig. 3D), we predict an intermediate state (FC) in which both the FB and CM regulators are turned on. Experiment on the mouse embryo cardio fibroblast cells (MECF) [25] found very similar expression pattern (Fig. 4D) to the FC state, supporting the existence of such intermediate state (FC) in the transdifferentiation process. We thus propose an experiment tracking both FB and CM regulators to further confirm our predictions. Table 1 summarizes a complete set of experimental agreements with the predicted cell states (see more experimental support in SI Section 4).

The ESN predictions are supported by not only the empirically observed cell states, but more importantly, the major reprogramming pathways (Figs. 2&3). For example, previous experiments found that the fibroblast-to-iPSC reprogramming pathways are very likely to undergo the pre-iPSC (PP) state [27,28]. Moreover, a major predicted pathway from FBs-to-CMs in transdifferentiation passing through the double positive state FC (both regulators of FBs and CMs expressed) is partially confirmed by previous works [27], and more experiments in future are needed to complete the validation. Table 2 summarizes the predicted major pathways along with its supporting experimental results.

**Methods**

***ODE model*** - We use following ODE model to describe the reprogramming dynamics

$$d\boldsymbol{x}/dt = \boldsymbol{F}(\boldsymbol{x}, \boldsymbol{\lambda}, \boldsymbol{\zeta}) + \boldsymbol{\eta}(\boldsymbol{x},t), \qquad (1)$$

where each component of $x$ represents the expression level of each gene, the term $F$ describes the gene interactions ($\lambda$ and $\zeta$ reflect chromatin open/close status and other intrinsic and environmental control parameters, respectively), and the stochasticity term $\eta$ satisfies $<\eta_i> = 0$ and $<\eta_i(t)\eta_j(t')> = 2D_{ij}\,\delta(t-t')$, where the matrix $D$ characterizes the strength of the stochastic noise (see SI Section 1).

***Transition Rate*** - The transition rate $k_{ab}$ estimated by the Wentzell-Freidlin theory for sufficiently small noises (SI Section 5, Fig. S11) is

$$k_{ab} \sim \exp\left(-\min\left(\int_{z_a}^{z_b} \sum_i \left(\frac{1}{2} D_{ii}^{-1}(dx_i/dt - F_i)\right) dx_i\right)\right), \tag{2}$$

where the integration is performed over the optimal path that minimizes the integral. Therefore Eq. (1) can be approximated by a master equation describing the network dynamics

$$dz/dt = K \cdot z, \tag{3}$$

where the $i$-th element of $z$ is the probability of finding the system in epigenetic state $i$, and $K = \{k_{ab}\}$ is the transition matrix determined by the edge weights.

***ESN construction*** - We construct the ESNs from ODE models by searching for both fixed-point attractors and first-order saddles through a conditional root-finding algorithm (SI Section 6). To achieve ensemble space appropriate for all chromatin statuses considered we use the following boundary conditions: For PR-ESN we set the chromatin region of CM regulators to be closed and other chromatin parts to be open (i.e., we set $\lambda = 0.1$ for CMR, and $\lambda = 1$ for other fully open chromatin parts). Moreover, both the FB and iPSC states have to occur within one single connected ESN cluster based on empirical observations; similarly, for TD-ESN we set $\lambda = 0.1$ for ESC regulators, and both the FB and CM states have to appear within a single connected ESN cluster; For GO-ESN, we combine sets of parameters from both PR-ESN and TD-ESN (see more details in SI Section 3).

**Acknowledgement** We are grateful to John Tyson, Fuchou Tang and Kathy Chen for illuminating constructive discussions. This work is supported by NSF DMS-0969417 (to J. X.). C. S. thanks NSF, ARL and ONR for support.

**Figure Captions**

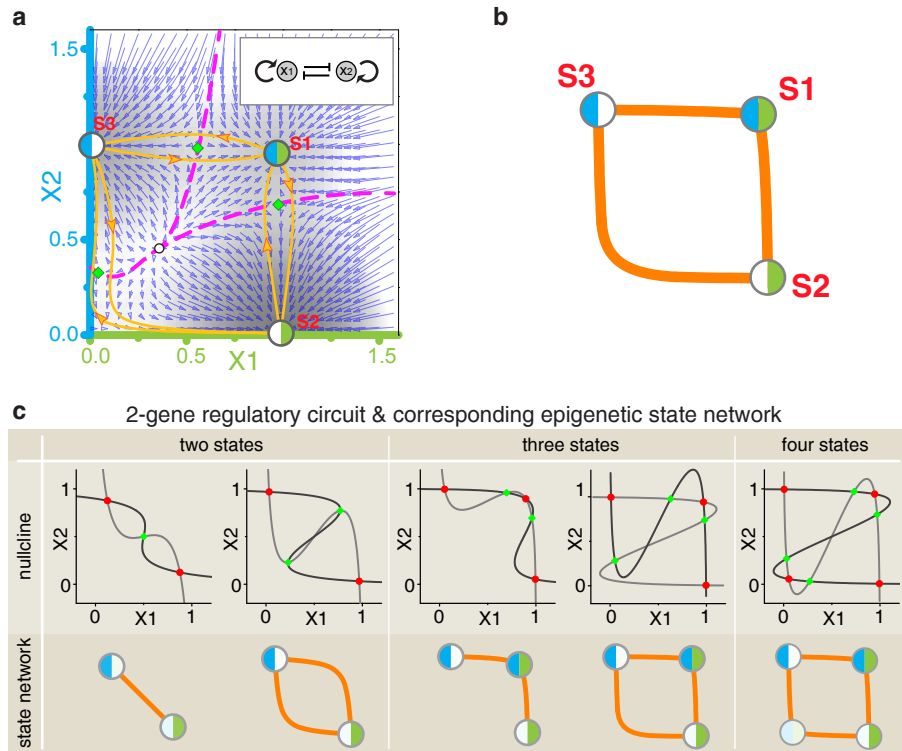

figure 1: A mutually inhibitive 2-gene regulatory circuit can result in different topologies of epigenetic state network. (A) The vector field shows fixed-point attractors (red points) separated by separatrices (pink lines) containing the first-order saddles (green points). Orange lines schematically show forward and backward optimal paths connecting two attractors, which actually bypass the saddle points as a signature of an open system (see Fig. S3). The insert is the regulatory circuit. (B) The epigenetic state network (ESN) corresponding to panel A. (C) Nullclines and the corresponding ESNs with different model parameters. More results and discussions are given in Table S1 and S2. In the figures, fixed-point attractors, separatrices, first-order saddle points, and the combined forward and backward optimal paths are represented by red points, pink lines, bright green points, and orange lines, respectively. The pie diagram of each node of an ESN represents the expression pattern of the corresponding state, with the left blue semi-circle representing $x_1$,

the right green semi-circle representing $x_2$, and the color depths reflecting the expression levels.

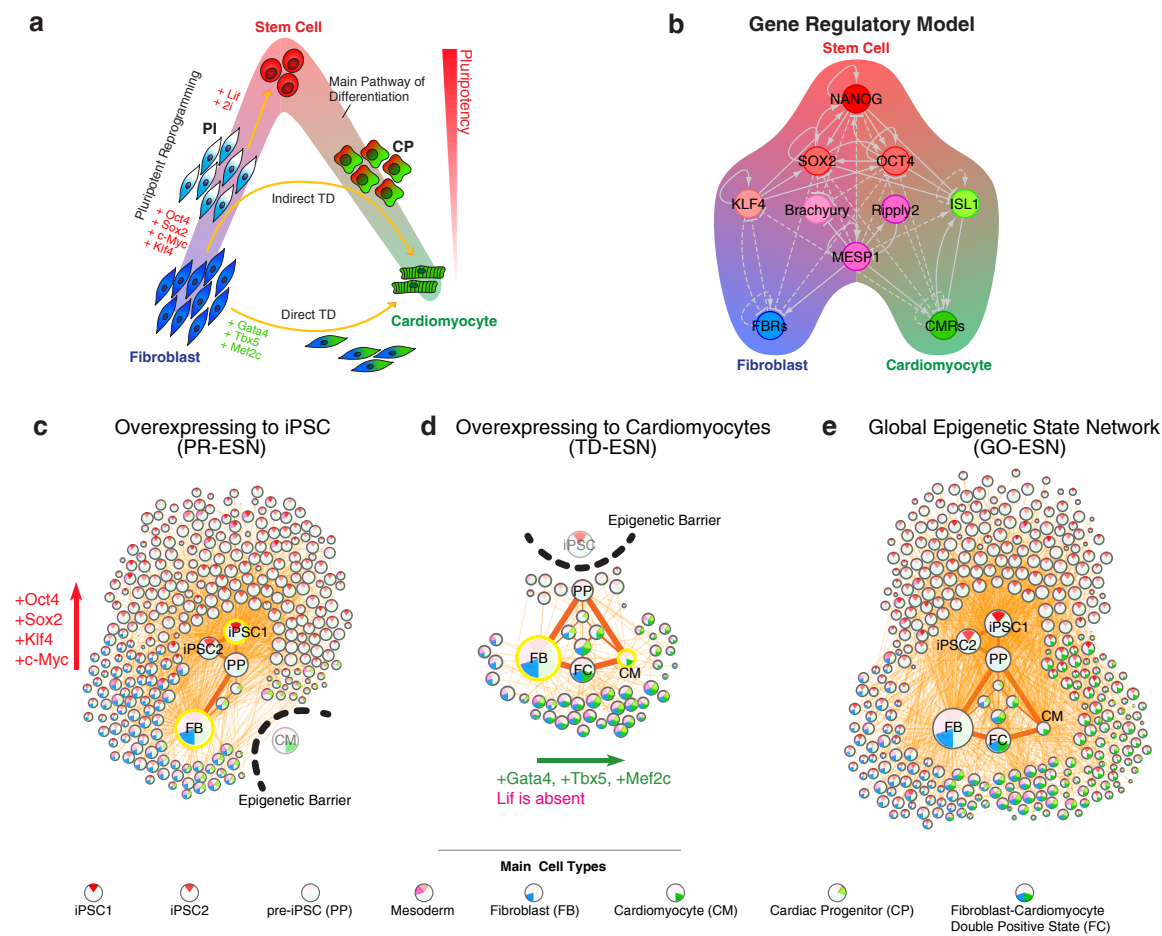

Figure 2: (A) Pluripotent reprogramming (PR) and transdifferentiation (TD) within the Fibroblast-iPSC-Cardiomyocyte system. Fibroblasts (FBs) can be induced to reprogram into pluripotent stem cells (SCs) then differentiate into cardiomyocytes (CMs), or to transdifferentiate into CMs through a direct or indirect pathway. (B) The Fibroblast-iPSC-Cardiomyocyte network constructed from literatures and the IPA® (@Ingenuity) database. See SOM5 for more details. (C) The calculated ensemble averaged ESN corresponding to the FB→iPSC pluripotent reprogramming process, total 295 states and 1014 edges obtained, with only existence of states FB and iPSC1 pre-required as experimental constraints. (differing in the relative expression levels of the stem cell regulators) (D) The calculated ensemble averaged ESN corresponding to the FB→CM transdifferentiation process, total 57 States 133 edges obtained, with only existence of states FB and CM pre-required. (E) The predicted ensemble averaged ESN for a global regulatory network, total 354 States and 1383 edges obtained. The three ESNs are obtained using the same sets of parameter, except existences of epigenetic barriers as indicated in Fig. S5. The node sizes are weighted by the averaged steady state probabilities, and the edge widths are weighted by the number of samples. Similar to Fig.1, the pie diagram of each node represents the expression pattern of the corresponding state, with the position and color of each slice corresponding to each group of regulators, starting at the central 12 clock position following the clockwise order, SC (red), Cardiac progenitor (green), CM (dark green), FB (blue), and Mesoderm (pink), and the color depth reflecting their average expression level. The yellow circles enclosing some pie diagrams indicate the pre-required states (see details in Table S3).

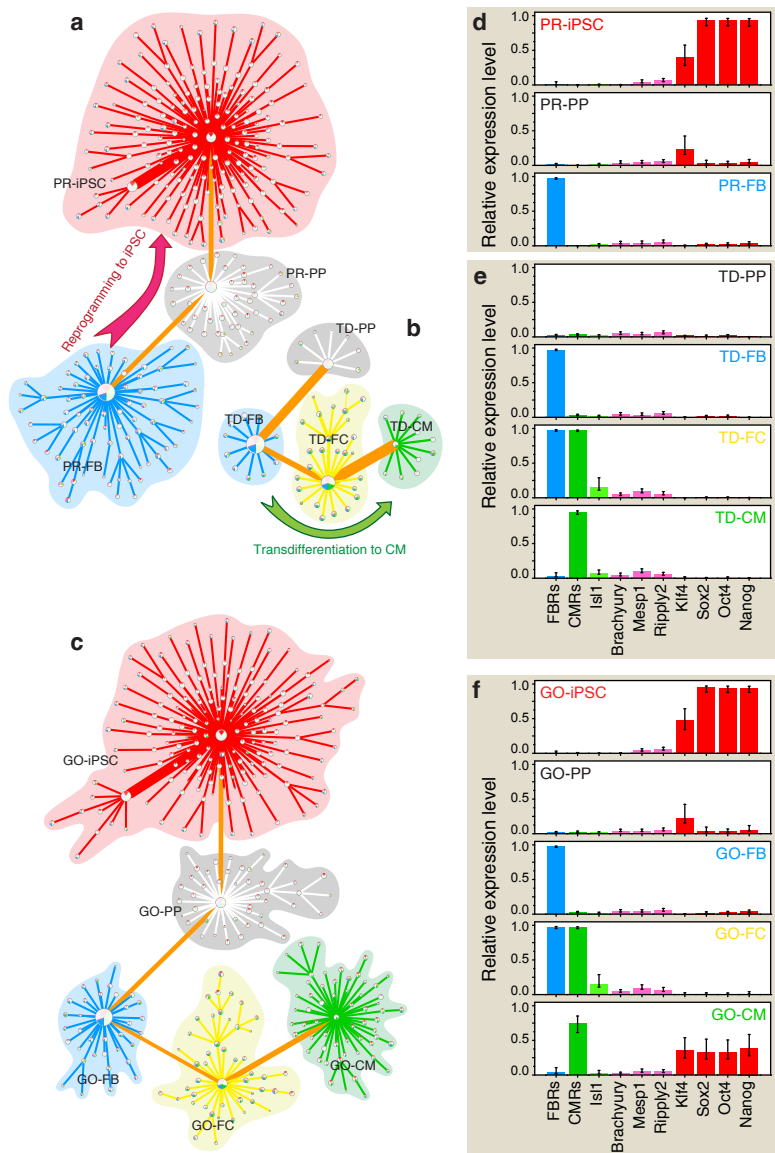

Figure 3: Minimum spanning trees of the (A) the FB→iPSC pluripotent reprogramming epigenetic state network (PR-ESN), (B) the FB→CM transdifferentiation epigenetic state network (TD-ESN), (C) the globally open epigenetic state network (GO-ESN), and (D-F) the predicted gene expression levels of the ten regulators averaged within each cluster. Same schemes as in Fig. 2 are used for the pie diagram representation of the states, node

and edge width and color. The error bars in D-F indicate expression level fluctuations within each cluster weighted by nodes' weights.

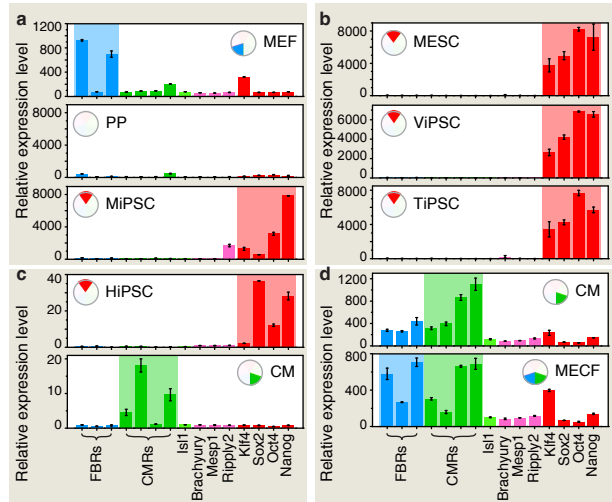

Figure 4: Measured relative expression levels of 15 key regulators from four experiments on cells undergoing different states during reprogramming and differentiation process. Twist1, Smail1 and Smail2 correspond to the fibroblast regulators (FBRs), and Nkx2-5, Tbx5, Gata4 and Mef2c correspond to the cardiomyocyte regulators (CMRs) in Fig. 3. (A) Cell states during mouse embryonic fibroblast (MEF) reprogramming to iPSC. Upper panel is 0 day post induction (dpi) corresponding to MEF state, middle panel is 3 dpi corresponding to pre-iPSC (PP) state, and the last panel is the final mouse iPSC (MiPSC) state. Data is re-plotted from GSE19023. (B) Comparison of mouse embryonic stem cell (MESC), ventricular cardiomyocyte induced pluripotent stem cell (ViPSC), and mouse tail-tip fibroblast induced pluripotent stem cell (TiPSC) shows heterogeneous stem cell patterns. Data is re-plotted from GSE32598. (C) Cell states of human induced pluripotent stem cell (HiPSC) differentiation to cardiomyocyte (CM). Data is re-plotted from GSE28191. (D) Comparison of mouse embryo cardiomyocyte (CM) and mouse embryo cardio fibroblast cells (MECF). Data is re-plotted from GSE14414.

Table 1: Experimental supports of cell states

| Cell State | Descriptions | Experimental Supports |
|---|---|---|
|  | Induced pluripotent stem cell (iPSC). | GEO accession GSE28191; GEO accession GSE19023 (J.-C. D. Heng *et al.*, 2011); GEO accession GSE32598 (H. Xu *et al.*, 2011). See Fig. 4. |
|  | Mesoderm cell (ME). | GEO accession GSE28191. See Fig. S7**D**. |
|  | Fibroblast (FB); Mouse embronic fibroblast (MEF); Mouse tail tip fibroblast (TTF). | GEO accession GSE19023 (J.-C. D. Heng *et al.*, 2011); GEO accession GSE32598 (H. Xu *et al.*, 2011). See Fig. 4**A**, S7**B**. |
|  | Cardiomyocyte (CM); Ventricular myocyte (VM). | GEO accession GSE28191; GEO accession GSE14414 (M. Ieda *et al.*, 2009); GEO accession GSE32598 (H. Xu *et al.*, 2011). See Fig. 4**C**, 4**D**, S7**B**. |
|  | Cardiac progenitor (CP). | GEO accession GSE28191. See Fig. S7**D**. |
|  | pre-iPSC (PP). Partiallly reprog-rammed cells which have not induced the endogenous pluripotent gene expression and but silenced somatic gene expression. | K. Plath *et al.*, 2011, and references therein on pre-iPSC; GEO accession GSE19023 (J.-C. D. Heng *et al.*, 2011). See Fig. 4**A**. |
|  | Fibroblast-Cardiomyocyte double positive cell (FC); Mouse embronic cardiac fibroblast (MECF). | GEO accession GSE14414 (M. Ieda *et al.*, 2009). See Fig. 4**C**. |
|  | Sub-phenotypes of iPSC with different epigenetic memories. | F. Soldner *et al.*, 2009; J. M. Polo *et al.*, 2010; H. Xu *et al.*, 2011. |

Table 1: List of experimental supports of predicted cell states.

Table 2: Experimental supports of reprogramming/transdifferientiation pathways

| Pathway | Descriptions | Experimental Supports |
|---|---|---|
| 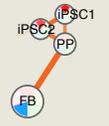 | Major pathway of reprgramming fibroblast to iPSC by passing through pre-iPSC state. | K. Takahashi *et al.*, 2007; K. Plath *et al.*, 2011, and references therein on pre-iPSC. |
| 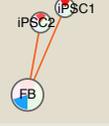 | Low-efficiency pathway from fibroblast to iPSC avoiding passing through or becoming trapped in the pre-iPSC cell stage. | S. Eminli *et al.*, 2008; J. Silva *et al*, 2008. |
| 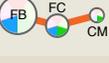 | Direct transdifferentiation from fibroblast to cardiomyocyte without passing through cardiac progenitor. | M. Ieda *et al.*, 2010. |
| 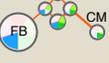 | Indirect transdifferentiation from fibroblast to cardiomyocyte through cardiac progenitor. | J. A. Efe *et al.*, 2011. |

Table 2: List of experimental supports of predicted transition pathways of reprogramming and transdifferentiation.